\documentstyle[aps,manuscript,psfig]{revtex}
\begin{document}
\draft
\renewcommand{\thefootnote}{\fnsymbol{footnote}}
\begin{title}
{\bf Hydrodynamic slip boundary condition at chemically patterned surfaces:
A continuum deduction from molecular dynamics}
\end{title}
\author{Tiezheng Qian and Xiao-Ping Wang}
\address{Department of Mathematics,
Hong Kong University of Science and Technology,\\
Clear Water Bay, Kowloon, Hong Kong, China} 
\author{Ping Sheng}
\address{Department of Physics and Institute of Nano Science and Technology,\\ 
Hong Kong University of Science and Technology,
Clear Water Bay, Kowloon, Hong Kong, China} 
\maketitle

\begin{abstract}
We investigate the slip boundary condition for single-phase flow 
past a chemically patterned surface. Molecular dynamics (MD)
simulations show that modulation of fluid-solid interaction along 
a chemically patterned surface induces a lateral structure 
in the fluid molecular organization near the surface. Consequently,
various forces and stresses in the fluid vary along the patterned surface.
Given the presence of these lateral variations, a general scheme is 
developed to extract hydrodynamic information from MD data.
With the help of this scheme, the validity of the Navier slip 
boundary condition is verified for the chemically patterned surface, 
where a local slip length can be defined. Based on the MD results,
a continuum hydrodynamic model is formulated using 
the Navier-Stokes equation and the Navier boundary condition, 
with a slip length varying along the patterned surface. 
Steady-state velocity fields from continuum calculations are
in quantitative agreement with those from MD simulations.
It is shown that, when the pattern period 
is sufficiently small, the solid surface appears to be homogeneous, 
with an effective slip length that can be controlled by surface patterning. 
Such a tunable slip length may have important applications in nanofluidics.
\end{abstract}
\narrowtext

\section{INTRODUCTION}\label{intro}

Recently developed techniques of microfluidics \cite{microfluidics} 
have generated great interest in further miniaturization towards
nanofluidics.
Precise control of fluid flow at nanoscale not only challenges 
existing techniques but also requires more realistic simulations 
for understanding nanoscale flow phenomena \cite{nanofluidics}.
Theoretical and experimental studies of confined fluids have shown 
that boundary conditions at various interfaces can strongly influence
the flow behavior as the system size approaches nanoscale. 
One particular boundary condition that has received substantial efforts
is the slip boundary condition at the fluid-solid interface 
\cite{nature-slip}. Molecular dynamics (MD) 
studies of fluid slipping have focused on single-phase flow past
homogeneous surfaces \cite{th-rob,th-tro,barrat,ckb,quirke-slip}
and miscible \cite{miscible-slip1,miscible-slip2} and 
immiscible \cite{immiscible-slip1,immiscible-slip2,qws-prl}
two-phase flows past homogeneous surfaces.
On the other hand, chemically patterned surfaces have generated
interest in novel control of flow in micro- and nanofluidics
\cite{CPS-flow0,CPS-flow1,CPS-flow2}.
In particular, slip flow past chemically patterned surfaces
has been investigated using both MD and continuum simulations
\cite{troian-preprint}.
Wetting and dewetting on chemically patterned surfaces have also been
extensively studied \cite{CPS-wetting1,CPS-wetting2}.

In this paper we study the slip boundary condition for single-phase
flow past a chemically patterned surface. 
MD simulations show that modulation of fluid-solid interaction along 
the patterned surface induces a lateral structure in the fluid 
molecular organization near the surface. As a consequence, various 
forces and stresses in the fluid vary along the patterned surface.
Given the presence of these lateral variations, a general scheme is 
developed to extract hydrodynamic information from MD data.
With the help of this scheme, the validity of the Navier slip 
boundary condition is verified for the chemically patterned surface, 
where a local slip length can be defined.
Continuum calculations are carried out using 
the Navier-Stokes equation and the Navier boundary condition (NBC),
with a slip length varying along the chemically patterned surface. 
Changing the ratio of the pattern period to 
the slip length, two distinct slip regimes are found. 
Particularly, when the pattern period is sufficiently small 
compared to the slip length, the solid surface appears to be approximately
homogeneous, with an effective slip length that can be controlled 
by surface patterning. Such a tunable slip length may have 
important applications in nanofluidics.

\begin{figure}[h]
\centerline{\psfig{figure=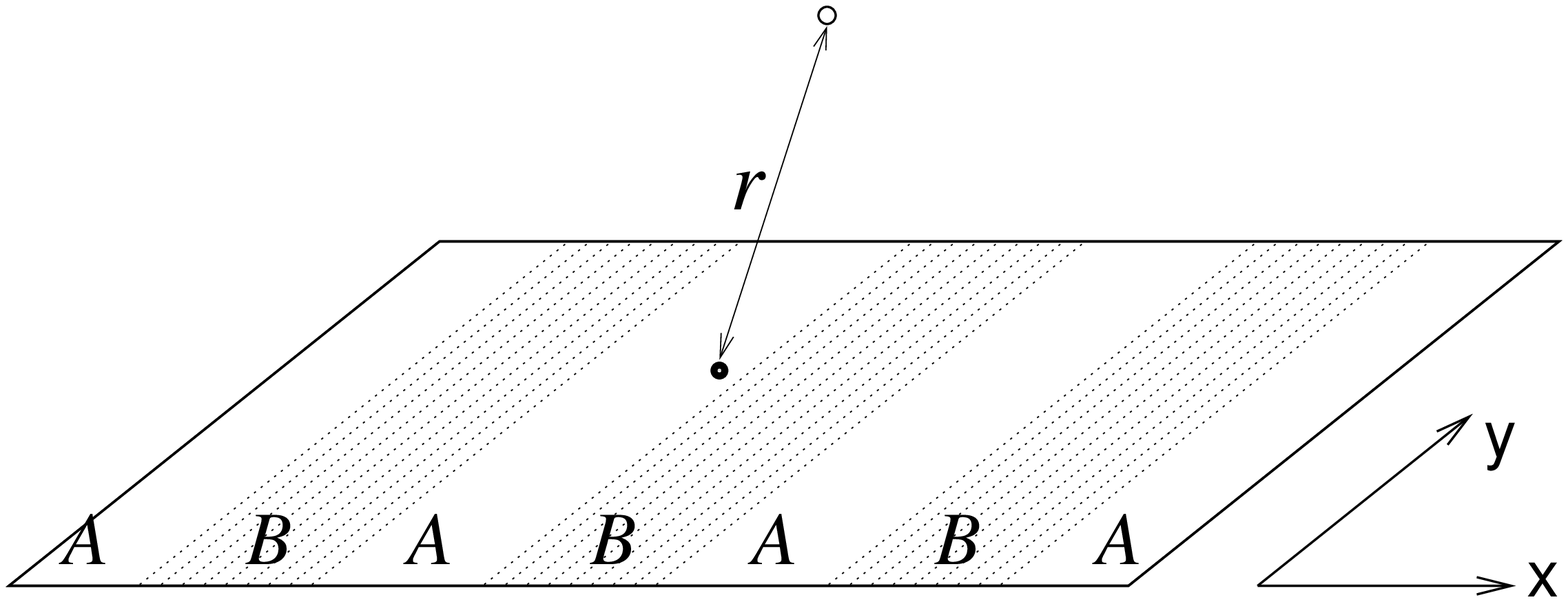,height=6.cm}}
\bigskip
\caption{\small Schematic of the solid planar surface which is
parallel to the $xy$ plane and patterned along the $x$ direction.
Interaction between a fluid molecule (empty circle) 
and a solid molecule (solid circle) separated by a distance $r$ 
is modeled by the potential $U_{fs}(r)$, with the parameter 
$\delta_{fs}$ varying with the equilibrium $x$ coordinate of 
the solid molecule.
}\label{fig_pattern}
\end{figure}

The paper is organized as follows. 
In Sec. \ref{md-tech}, we describe the MD simulation techniques 
for a single-phase fluid confined between two solid planar walls,
one with a chemically homogeneous surface and 
the other a chemically patterned surface.
The solid surfaces are parallel to the $xy$ plane 
and surface patterning is along the $x$ direction 
(see Fig. \ref{fig_pattern}).
Interaction between fluid and solid molecules is modeled by 
a modified Lennard-Jones (LJ) potential 
$$U_{fs}(r)=4\epsilon_{fs}\left[\left(\sigma_{fs}/r\right)^{12}-
\delta_{fs}\left(\sigma_{fs}/r\right)^6\right],$$
where $\epsilon_{fs}$ and $\sigma_{fs}$ are the energy and range
parameters, and $\delta_{fs}$ is a dimensionless parameter 
adjusting the wetting property of the fluid.
This LJ potential has constant parameters $\epsilon_{fs}$, 
$\sigma_{fs}$, and $\delta_{fs}$  at the homogeneous surface.
For the patterned surface, we choose $\delta_{fs}$ to be
an interaction parameter that varies with the equilibrium 
$x$ coordinate of the solid molecule (see Fig. \ref{fig_pattern}).
Denoted by $\delta_{fs}(x)$, the variation of $\delta_{fs}$ 
in the $x$ direction models the patterned surface in MD simulations.

In Sec. \ref{static-dynamic}, we describe how to extract 
hydrodynamic information from MD data.
It is shown that in the absence of hydrodynamic motion, 
modulation of fluid-solid interaction along the patterned surface 
induces a nonuniform static state in which various forces 
and stresses in the fluid vary along the solid surface, and these 
lateral variations decay quickly away from the fluid-solid interface.
The forces and stresses measured in the static equilibrium state 
are taken as reference quantities.
Together they form a molecular background from which 
hydrodynamic components are generated when the fluid is driven into
hydrodynamic motion. Couette flow is generated by shearing the fluid.
The forces and stresses measured in the dynamic steady state 
are taken as dynamic quantities. A hydrodynamic quantity is 
obtained by subtracting the reference quantity 
from the corresponding dynamic quantity:
$$\tilde{Q}=Q_{D}-Q_{S},$$
where $Q_{S}$ is a static (reference) quantity,
$Q_{D}$ the corresponding dynamic quantity,
and the over tilde denotes the hydrodynamic quantity.

In Sec. \ref{model}, we formulate a continuum hydrodynamic model
for slip flow past chemically patterned surfaces.
Various MD data are collected and analyzed under the subtraction scheme
described in Sec. \ref{static-dynamic}. In particular,
the NBC is verified to be a local constitutive relation for 
the hydrodynamic fluid-solid coupling at the patterned surface.  
According to the NBC, the slip velocity is locally proportional to 
the hydrodynamic tangential stress at the wall, 
and the local proportionality constant defines a local slip length.
This may be expressed by
$$\beta(x)v_x^{slip}(x)=-\tilde{\sigma}_{nx}=-\eta \partial_n v_x(x),$$
where $v_x^{slip}$ is the slip velocity, $-\tilde{\sigma}_{nx}=
-\eta \partial_n v_x(x)$ the hydrodynamic tangential viscous stress 
($\eta$ is the viscosity and $n$ denotes the outward surface normal), 
and $\beta(x)$ the local slip coefficient. 
It is shown that as a local property at the patterned surface,
$\beta(x)$ is determined by the local interaction parameter
$\delta_{fs}(x)$, as expressed by $\beta(x)=\beta(\delta_{fs}(x))$,
where $\beta$ as a function of $\delta_{fs}$,
denoted by $\beta(\delta_{fs})$, relates a microscopic interaction
parameter and a coefficient in continuum boundary condition.
The local slip length $l_s(x)$, defined by 
$l_s(x)=\eta/\beta(\delta_{fs}(x))$, varies with 
the local fluid-solid interaction at the patterned surface.

The continuum hydrodynamic model formulated in Sec. \ref{model} 
consists of the Navier-Stokes equation with a constant viscosity 
and the NBC with a laterally varying slip length.
All the parameters involved in the continuum model calculations
are directly determined from MD simulations, 
thus no adjustable parameter appears.
In Sec. \ref{continuum}, we present steady-state velocity fields 
from continuum calculations,
in quantitative agreement with those from MD simulations.
This further affirms the validity of the local NBC.
Continuum calculations also show that there are two distinct regimes 
for fluid slipping.
As the surface pattern period is much larger than the slip length,
well separated slip domains form along the solid surface. 
When the pattern period is sufficiently small, by contrast,
the solid surface appears to be homogeneous and 
the slip velocity can be predicted using an effective slip length.
The latter regime provides a way to systematically tune
the slip length. We conclude this paper in Sec. \ref{disc}.

\section{MOLECULAR DYNAMICS SIMULATIONS}\label{md-tech}

\subsection{Geometry and Interactions}

\begin{figure}[ht]
\centerline{\psfig{figure=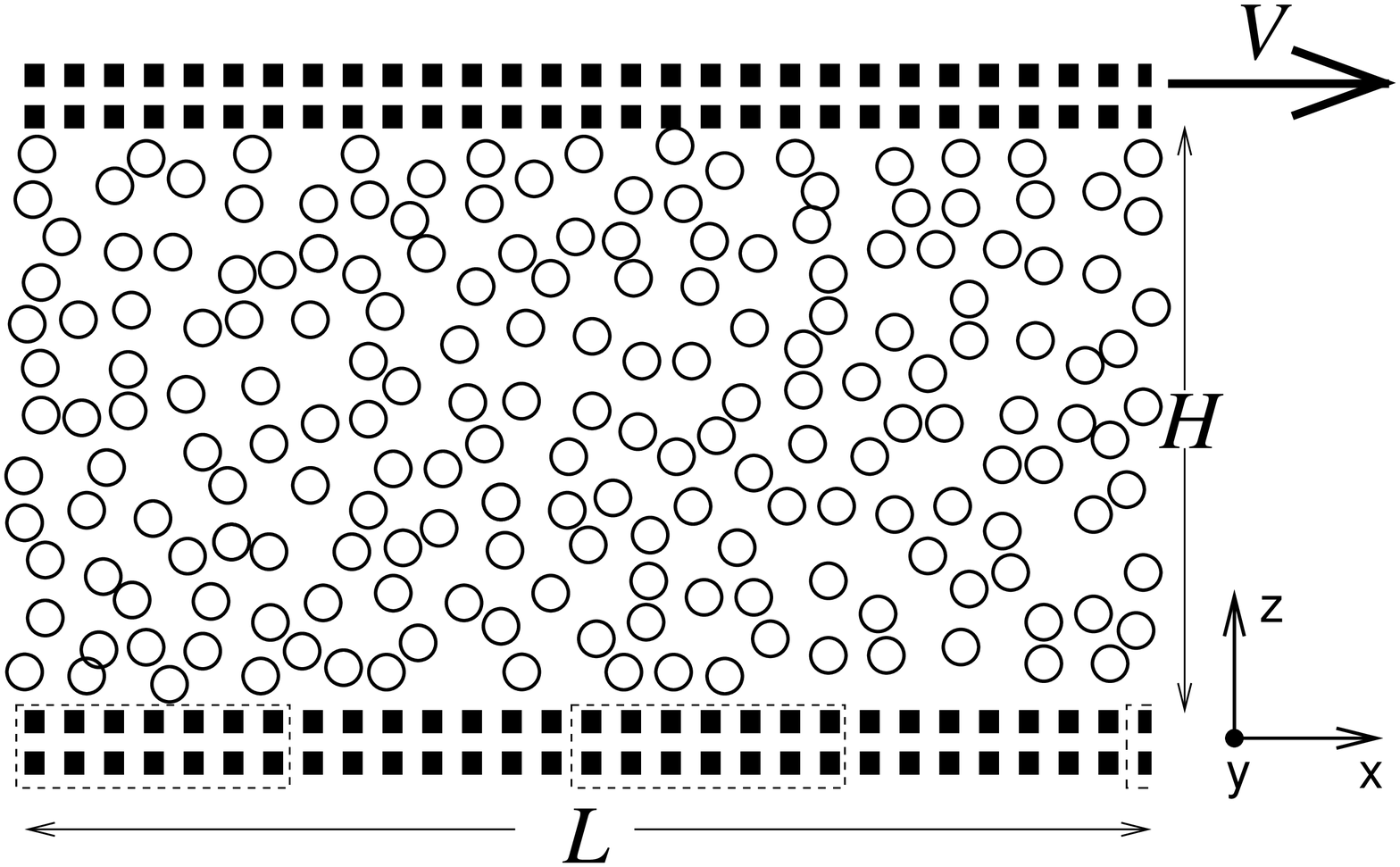,height=10cm}}
\bigskip
\caption{\small Schematic of the Couette-flow geometry adopted 
in MD simulations. The lower solid surface is patterned 
in the $x$ direction and the upper surface is homogeneous. 
Moving the upper wall with speed $V$ along $x$ generates 
the Couette flow. 
}\label{fig_geometry}
\end{figure}

\begin{figure}[h]
\centerline{\psfig{figure=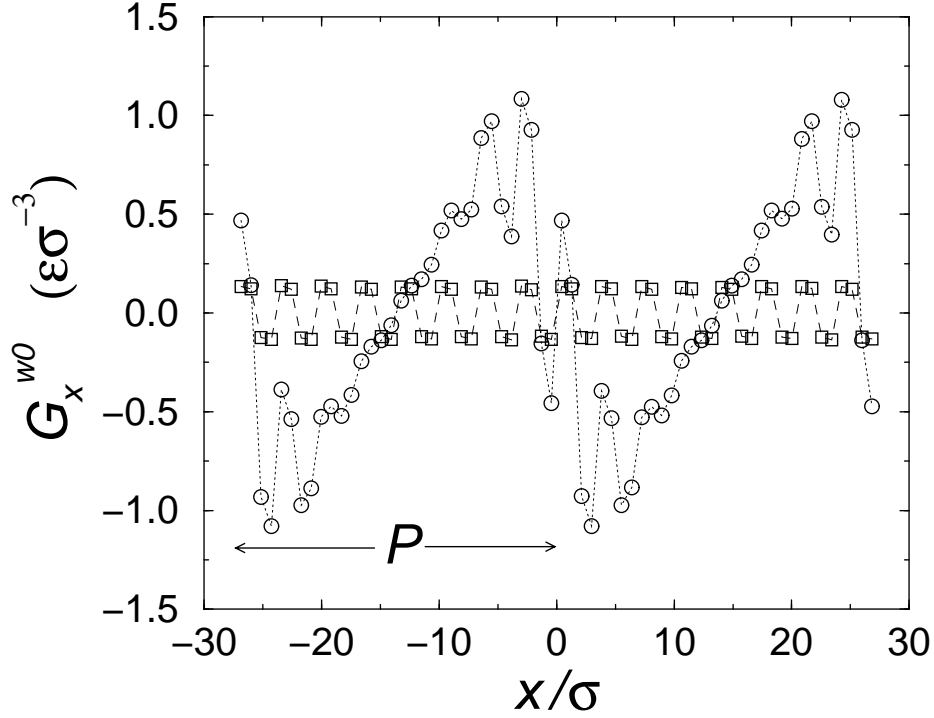,height=10cm}}
\bigskip
\caption{\small Static tangential wall force $G_x^{w0}$ 
plotted as a function of $x$.
The circles denote $G_x^{w0}=\int_0^{\Delta z} dzg_x^{w0}(x,z)$ 
measured in the boundary layer at the lower fluid-solid interface, 
showing a fast oscillation superimposed on a slow variation of period $P$; 
the squares denote $G_x^{w0}=\int_{H-\Delta z}^H dzg_x^{w0}(x,z)$ 
measured in the boundary layer at the upper interface, 
showing a fast oscillation only. 
Here $g_x^{w0}$ is the tangential wall force density.
}\label{fig_wallforce0}
\end{figure}

\begin{figure}[h]
\centerline{\psfig{figure=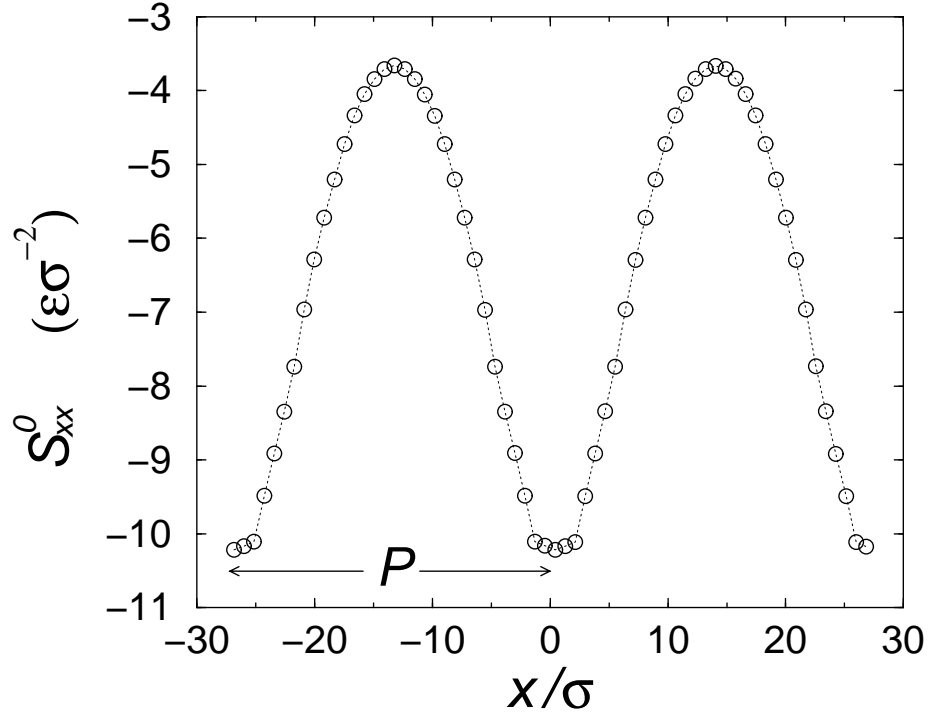,height=10cm}}
\bigskip
\caption{\small $z$-integrated static normal stress across 
the boundary layer, $S^0_{xx}=
\int_0^{\Delta z}dz\sigma_{xx}^0(x,z)$, plotted as a function of $x$.
}\label{fig_normalstress0}
\end{figure}

\begin{figure}[h]
\centerline{\psfig{figure=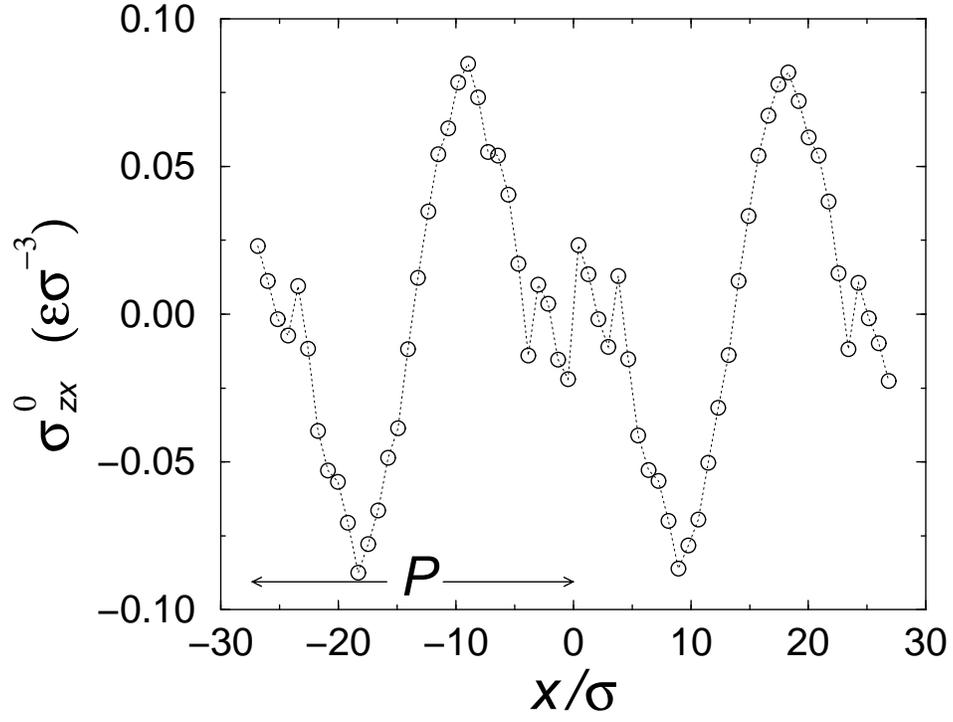,height=10cm}}
\bigskip
\caption{\small Static tangential stress $\sigma_{zx}^0(x,z)$ 
at $z=\Delta z$, plotted as a function of $x$.
}\label{fig_tangentialstress0}
\end{figure}

\begin{figure}[h]
\centerline{\psfig{figure=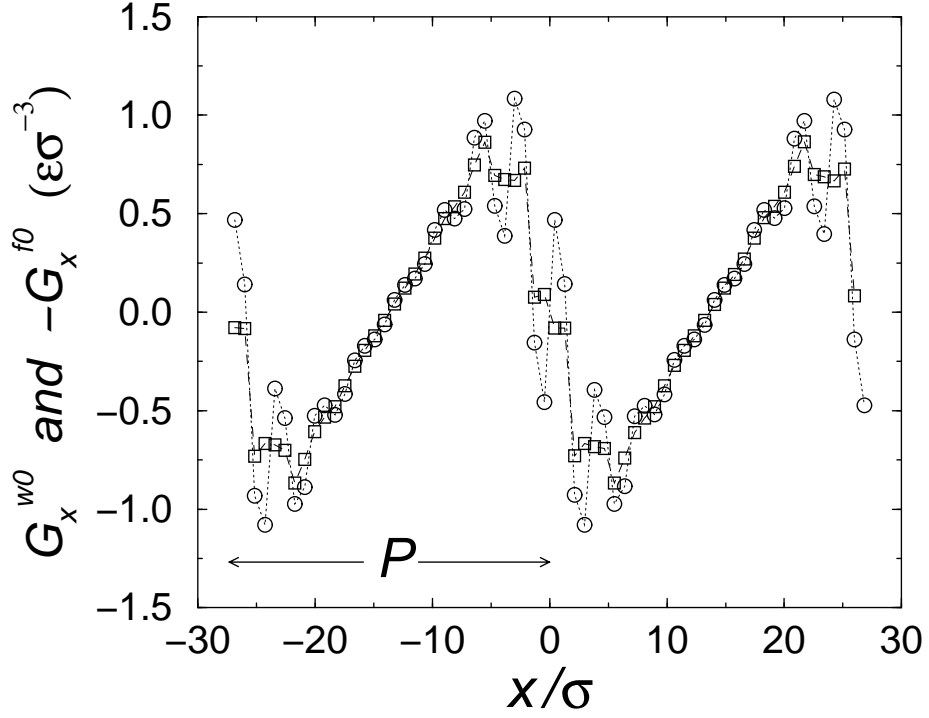,height=10cm}}
\bigskip
\caption{\small Static tangential wall force ${G}_x^{w0}(x,\Delta z/2)$ 
and the negative of static tangential fluid force $-{G}_x^{f0}(x,\Delta z/2)$ 
measured in the boundary layer at the lower fluid-solid interface. 
The circles denote ${G}_x^{w0}$ (see Fig. \ref{fig_wallforce0});
the squares denote $-{G}_x^{f0}$. 
It is seen that ${G}_x^{w0}+{G}_x^{f0}=0$ (Eq. (\ref{BL-force-balance0})).
Our stress measurement scheme has smoothed the kinetic contribution to
$\sigma_{xx}^0$. That's why the fast oscillation in $-{G}_x^{f0}$ 
is less prominent than that in ${G}_x^{w0}$.
}\label{fig_forcebalance0}
\end{figure}

\begin{figure}[h]
\centerline{\psfig{figure=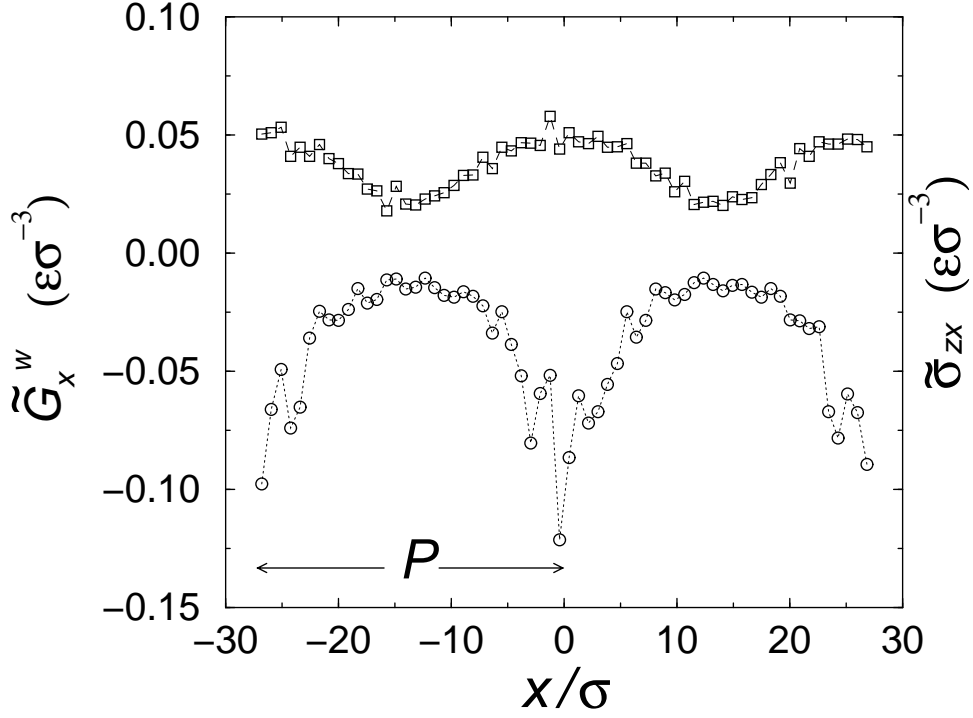,height=10cm}}
\bigskip
\caption{\small Hydrodynamic tangential wall force
$\tilde{G}_x^{w}(x,\Delta z/2)$ (circles) and stress
$\tilde{\sigma}_{zx}(x,\Delta z)$ (squares) for the boundary layer at 
the lower fluid-solid interface. Note that the magnitude for the variation 
in each hydrodynamic quantity is much smaller than that seen from
the corresponding static quantity (see Figs. \ref{fig_wallforce0}
and \ref{fig_tangentialstress0}).
}\label{fig_BLforceshyd1}
\end{figure}

\begin{figure}[h]
\centerline{\psfig{figure=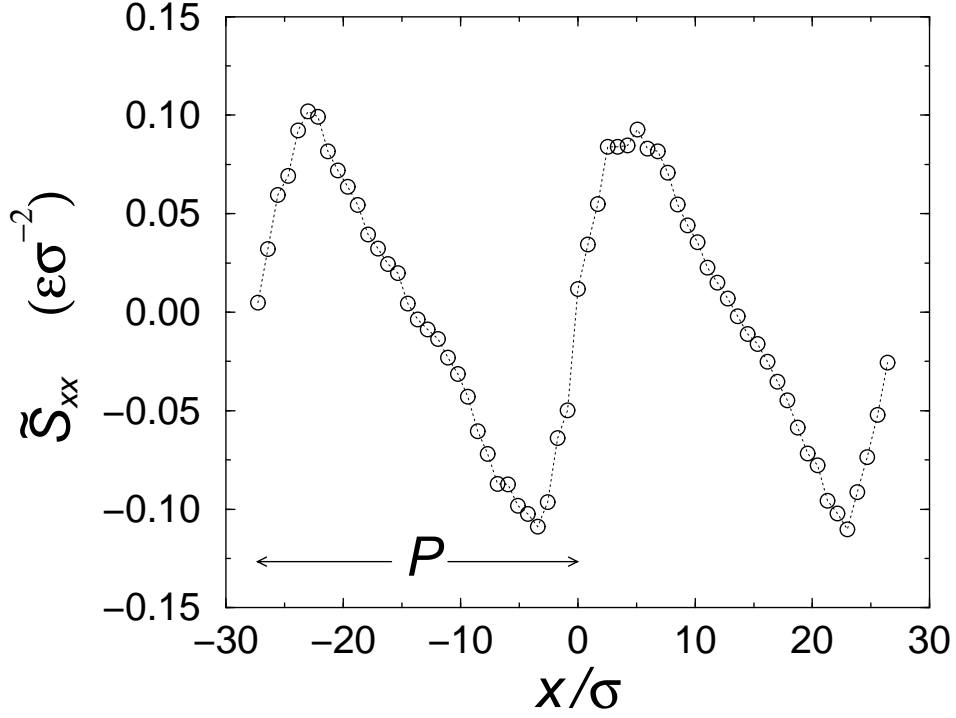,height=10cm}}
\bigskip
\caption{\small $z$-integrated hydrodynamic normal stress across 
the boundary layer at the lower fluid-solid interface, 
$\tilde{S}^0_{xx}=\int_0^{\Delta z}dz\tilde{\sigma}_{xx}^0(x,z)$, 
plotted as a function of $x$.
Note that the magnitude for the variation in this hydrodynamic quantity 
is much smaller than that seen from the corresponding static quantity 
(see Fig. \ref{fig_normalstress0}).
}\label{fig_BLforceshyd2}
\end{figure}

\begin{figure}[h]
\centerline{\psfig{figure=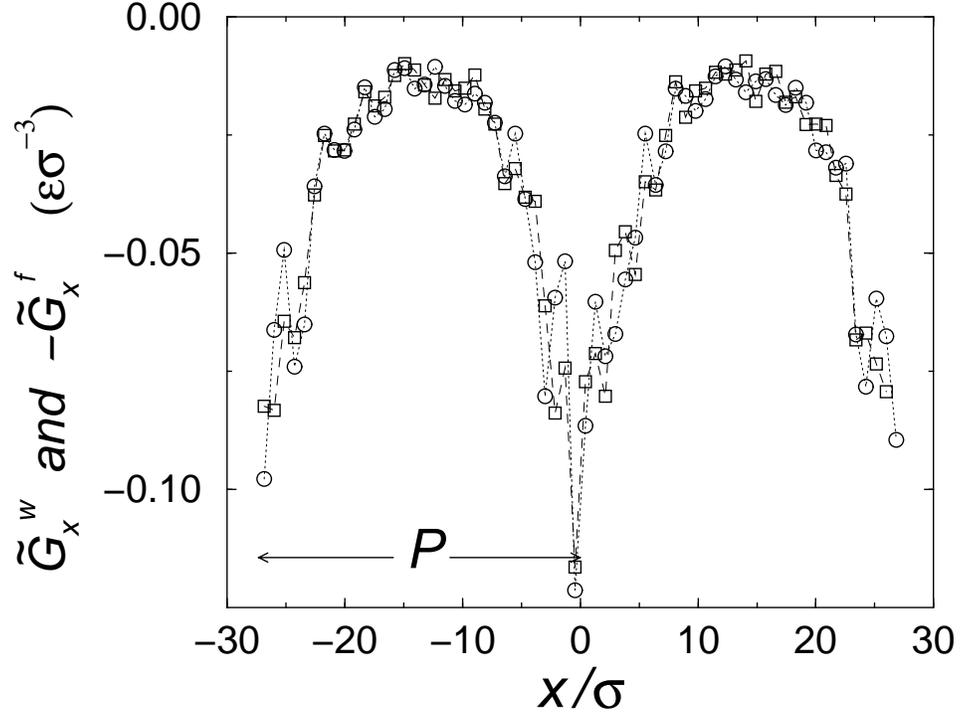,height=10cm}}
\bigskip
\caption{\small Hydrodynamic tangential wall force 
$\tilde{G}_x^{w}(x,\Delta z/2)$ and the negative of hydrodynamic tangential
fluid force $-\tilde{G}_x^{f}(x,\Delta z/2)$ in the boundary layer at 
the lower fluid-solid interface. The circles denote $\tilde{G}_x^{w}$
(see Fig. \ref{fig_BLforceshyd1}); the squares denote $-\tilde{G}_x^{f}$. 
It is seen that $\tilde{G}_x^{w}+\tilde{G}_x^{f}=0$ 
(Eq. (\ref{BL-force-balancehyd})).
}\label{fig_forcebalancehyd}
\end{figure}

\begin{figure}[h]
\centerline{\psfig{figure=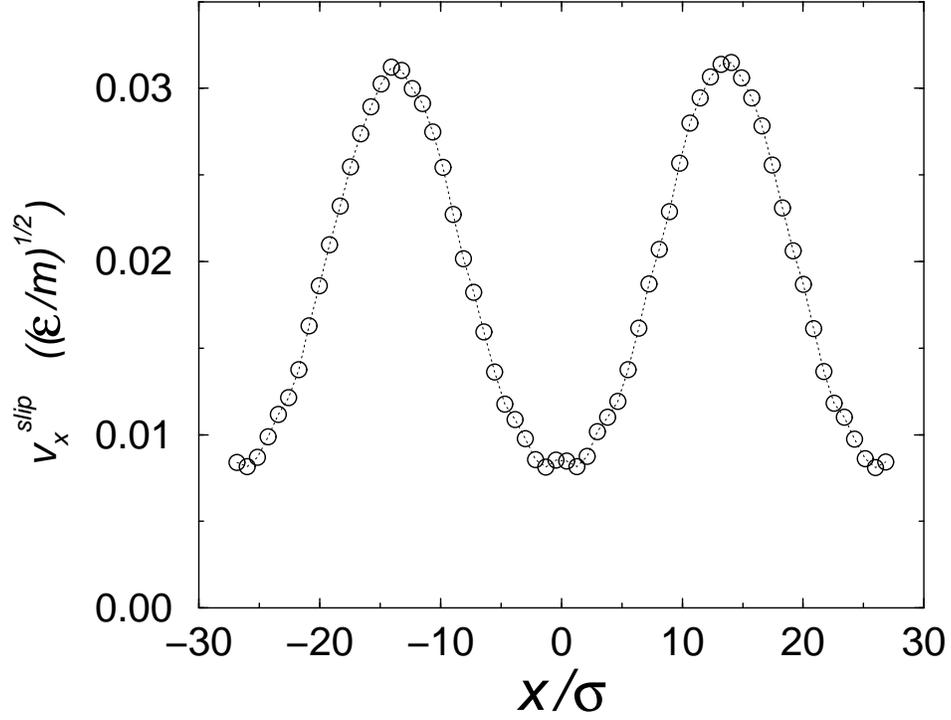,height=10cm}}
\bigskip
\caption{\small Slip velocity $v_x^{slip}(x)$ at 
the patterned surface, plotted as a function of $x$. 
It is seen that $v_x^{slip}$ varies periodically along $x$, 
being small in the large $\delta_{fs}$ (large $\beta$) region and 
large in the small $\delta_{fs}$ (small $\beta$) region.
Larger the $\beta$ coefficient, smaller the slip length 
and slip velocity. The no-slip boundary condition corresponds to 
the limit of $\beta\rightarrow\infty$.}
\label{fig_vxslip}
\end{figure}

\begin{figure}[h]
\centerline{\psfig{figure=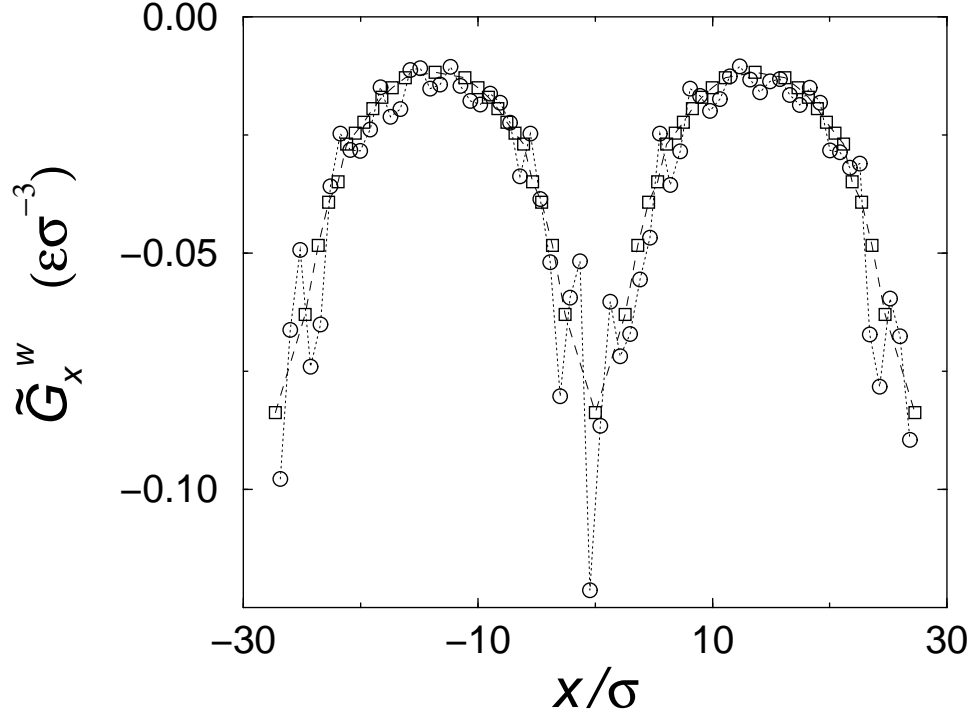,height=10cm}}
\bigskip
\caption{\small Hydrodynamic tangential wall force 
$\tilde{G}_x^{w}(x)$ at the lower fluid-solid interface. 
The circles denote $\tilde{G}_x^{w}(x)$ directly measured 
in the MD simulation (see Fig. \ref{fig_BLforceshyd1});
the squares denote $\tilde{G}_x^{w}(x)$ calculated from $\beta(x)$ 
and the measured $v_x^{slip}(x)$ according to Eq. (\ref{navier-wallforce}).
}\label{fig_wallforcecomparison}
\end{figure}

\begin{figure}[h]
\centerline{\psfig{figure=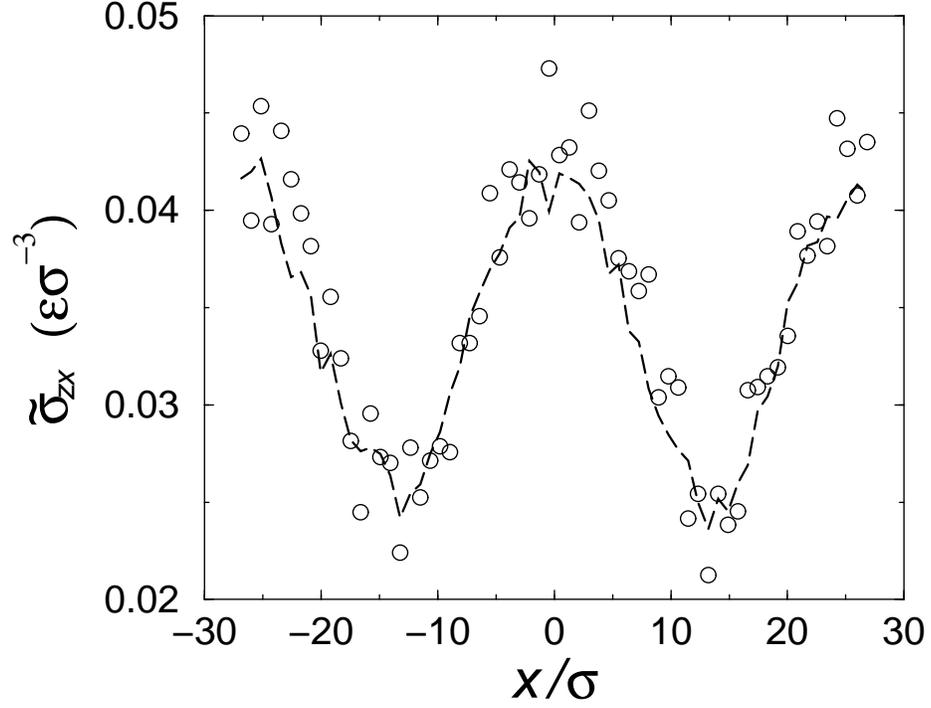,height=10cm}}
\bigskip
\caption{\small Hydrodynamic tangential stress 
$\tilde{\sigma}_{zx}(x,z)$ at $z=2\Delta z$, 
plotted as a function of $x$. The circles denote 
$\sigma_{zx}(x,2\Delta z)-\sigma_{zx}^0(x,2\Delta z)$,
obtained according to the subtraction scheme for MD data;
the dashed line denotes $\eta (\partial_zv_x+\partial_xv_z)$, 
calculated from the MD flow field.
Here $\eta=2.3\sqrt{\epsilon m}/\sigma^2$ has been used to achieve the
best agreement.
}\label{fig_newton}
\end{figure}

MD simulations have been carried out for Couette flow 
between two solid walls, one with a chemically homogeneous
surface and the other a chemically patterned surface  
(see Fig. \ref{fig_geometry}). 
The two planar walls are parallel to the $xy$ plane, 
with the two fluid-solid boundaries defined at $z=0$, $H$. 
The Couette flow is generated by moving the upper wall 
at a constant speed $V$ in the $x$ direction. Periodic boundary 
conditions are imposed in the $x$ and $y$ directions.
Interaction between fluid molecules separated by a distance $r$ 
is modeled by a Lennard-Jones (LJ) potential 
$U_{ff}(r)=4\epsilon\left[\left(\sigma/r\right)^{12}-
\left(\sigma/r\right)^6\right]$,
where $\epsilon$ is the energy scale and $\sigma$ the range scale.
The average number density for the fluid is set at 
$\rho=0.81\sigma^{-3}$. 
The temperature is maintained at $T=1.4\epsilon/k_B$. 
Each wall is constructed by two [001] planes of an fcc lattice, 
with each wall molecule attached to a lattice site by 
a harmonic spring. The mass of the wall molecule equals 
that of the fluid molecule $m$. 
The number density of the wall equals $\rho_w=1.86\sigma^{-3}$.
The fluid-solid interaction is modeled by a modified LJ potential 
$U_{fs}(r)=4\epsilon_{fs}\left[\left(\sigma_{fs}/r\right)^{12}-
\delta_{fs}\left(\sigma_{fs}/r\right)^6\right]$,
with the energy and range parameters given by 
$\epsilon_{fs}=1.16\epsilon$ and $\sigma_{fs}=1.04\sigma$, 
and a dimensionless parameter $\delta_{fs}$ 
for adjusting the wetting property of the fluid.
The lower solid surface is patterned, and the patterning effect  
is modeled by the oscillating parameter \cite{weiqing}
\begin{equation}\label{periodic-delta}
\delta_{fs}(x)=\delta_0+\delta_1\cos\left[ 
\displaystyle\frac{2\pi x}{P}\right],
\end{equation}
where $x$ is the $x$-coordinate of the wall molecule, 
$\delta_0$ is a constant,
$\delta_1$ is the oscillation amplitude for $\delta_{fs}$,
and $P$ is the oscillation period, set to be larger than $10\sigma$.
(This requirement will be explained in Sec. \ref{slip-coefficient}.)
The upper solid surface is homogeneous, at which 
the fluid-solid interaction has a constant $\delta_{fs}$, equaling $\delta_0$. 
We use $\delta_0=1$ and $\delta_1=0.3$. 
The fluid-solid interaction potential $U_{fs}$ is cut off at 
$r_c=2.2\sigma$.
(Thicker walls with larger cut-off distance have also been used, 
but the MD results are quite stable. 
See Appendix \ref{app-atomistic-solid-walls}.)
In most of our simulations, the shearing speed $V$ is on the order 
of $0.1\sqrt{\epsilon/m}$, the wall separation along $z$ is 
$H=13.6\sigma$, the sample dimension along $y$ is $6.8\sigma$,
and the sample dimension along $x$ is set to be $L=2P$ 
(from $-P$ to $P$). 
The MD results presented in Figs. \ref{fig_wallforce0} to
\ref{fig_forcebalancehyd} and \ref{fig_vxslip} to 
\ref{fig_vx_osci} are obtained for $V=0$ and 
$0.25\sqrt{\epsilon/m}$, $H=13.6\sigma$, and $L=2P=54.5\sigma$.

\subsection{Measurements}

To have spatial resolution along the $x$ and $z$ directions, 
the measurement region is evenly divided into bins, 
each $\Delta x=0.85\sigma$ by 
$\Delta z =0.85\sigma$ in size.
That is, the fluid space is divided along the $z$ direction 
into $H/\Delta z$ horizontal layers, and each layer is then divided 
along the $x$ direction into $L/\Delta x$ bins.
We measure five quantities: $G_x^w$, the tangential wall force 
per unit area exerted by the wall on the fluid molecules 
in a horizontal layer; 
$\sigma_{xx}$ and $\sigma_{zx}$, the $xx$ and $zx$ components of 
the fluid stress tensor; $v_x$ and $v_z$, the $x$ and $z$
components of the fluid velocity.
$G_x^w$ is obtained from the time average of 
the total tangential wall force experienced by the fluid molecules 
in a bin, divided by the bin area in the $xy$ plane 
(for more details see Appendix \ref{app-tangential-wall-force}); 
$\sigma_{xx(zx)}$ is obtained from the time average of 
the kinetic momentum transfer plus the fluid-fluid interaction forces
across a constant-$x(z)$ bin surface 
(for more details see Appendix \ref{app-stress-measurement}); 
$v_{x}$ and $v_z$ are measured from the time average of fluid molecules' 
velocities in each bin. As reference quantities, we also measure 
$G_x^{w0}$, $\sigma_{xx}^0$, and $\sigma_{zx}^0$  
in the static ($V=0$) state (in which $v_x$ and $v_z$ both vanish). 
Here the superscript ``0'' denotes the static quantities.
Static equilibrium-state ($V=0$) and dynamic steady-state ($V\ne 0$) 
quantities are obtained from time average over 
$5\times 10^5\tau$ or longer where $\tau$ is the atomic time scale 
$\sqrt{m\sigma^2/\epsilon}$. 
Throughout the remainder of this paper, all physical quantities 
are given in terms of the LJ reduced units (defined in terms of 
$\epsilon$, $\sigma$, and $m$).

\section{STATIC AND DYNAMIC STATES}\label{static-dynamic}

\subsection{Static Equilibrium State}

\subsubsection{Nonuniform static state}

In the present Couette flow geometry, a static equilibrium state 
is usually characterized by zero velocity and constant pressure 
in a continuum hydrodynamic description. 
Microscopically, however, various forces and stresses 
in the fluid actually vary along the solid surface.
These lateral variations are induced by the short-range 
fluid-solid interaction, and they decay quickly away from 
the fluid-solid interface. The periodicity of the surface structure 
necessarily shows up in these near-surface lateral variations. 
First there is the short period of crystalline structure. 
(For $G_x^{w0}$, this corresponds to the first component 
discussed in Appendix \ref{app-tangential-wall-force}.)
At the lower fluid-solid interface, in addition, there is 
the large period of surface patterning: the fluid-solid interaction 
itself is periodically modulated along the solid wall, 
with a period much larger than that of the crystalline structure. 
As a consequence, the lateral force/stress variations will show 
this period of the fluid-solid interaction.
(For $G_x^{w0}$, this corresponds to the second component discussed 
in Appendix \ref{app-tangential-wall-force}.)
In Fig. \ref{fig_wallforce0}, we show the profiles of $G_x^{w0}$ 
for the two boundary layers at the lower and upper fluid-solid interfaces. 
While $G_x^{w0}$ at the upper fluid-solid interface shows only 
a fast oscillation imposed by the crystalline structure, 
at the lower interface $G_x^{w0}$ shows a slow variation as well, 
induced by the periodicity of the fluid-solid interaction.
In Fig. \ref{fig_normalstress0}, we show the profile of
$z$-integrated $\sigma_{xx}^0$ for the boundary layer at the lower 
fluid-solid interface. A slow variation with period $P$ is clearly seen.
In Fig. \ref{fig_tangentialstress0}, we show the profile of 
$\sigma_{zx}^0$ at $z=\Delta z$ near the lower fluid-solid interface. 
Once again, a slow variation with period $P$ is clearly seen.

\subsubsection{Static force balance}

In the static equilibrium state ($V=0$), local force balance 
necessarily requires the stress tangential to the fluid-solid 
interface to be the same on the two sides. 
Then the wall force and fluid stress must vary in such a way 
that the total force density vanishes. 
An integrated form for the static tangential force balance 
in the boundary layer at the lower fluid-solid interface is given by
\begin{equation}\label{BL-force-balance0}
G_x^{w0}(x,\Delta z/2)+{G}_x^{f0}(x,\Delta z/2)=0, 
\end{equation}
Here $\Delta z/2$ represents the mid-level $z=\Delta z/2$
in the boundary layer between $z=0$ and $z=\Delta z$, 
$G_x^{w0}(x,\Delta z/2)$ is the tangential wall force per unit area 
in this layer, given by
$$G_x^{w0}(x,\Delta z/2)=\int_0^{\Delta z}dz g_x^{w0}(x,z)$$ 
in terms of the tangential wall force density $g_x^{w0}$,
and $G_x^{f0}(x,\Delta z/2)$ is the tangential fluid force 
per unit area in the same layer, given by
\begin{equation}\label{fluid-force-expression0}
\begin{array}{ll}
{G}_x^{f0}(x,\Delta z/2) & =\int_0^{\Delta z}dz
[\partial_x{\sigma}_{xx}^0(x,z)+\partial_z{\sigma}_{zx}^0(x,z)]\\
& =\partial_x\int_0^{\Delta z}dz{\sigma}^0_{xx}(x,z)
+{\sigma}_{zx}^0(x,\Delta z),
\end{array}
\end{equation}
where ${\sigma}_{xx}^0$ and ${\sigma}_{zx}^0$ are the $xx$ and $zx$
components of fluid stress in the static state, both measured
as reference quantities. In Eq. (\ref{fluid-force-expression0}) 
the fact that $\sigma_{zx}^0(x,0)=0$ has been used. 
(More strictly, $\sigma_{zx}^0(x,0^-)=0$ because there is no fluid 
below $z=0$, hence no momentum transport across $z=0$.)
Similar expressions for the static tangential force balance in other
horizontal layers can be obtained as well.
In Fig. \ref{fig_forcebalance0} we show
the profiles of ${G}_x^{w0}$ and $-{G}_x^{f0}$ for the boundary layer 
at the lower fluid-solid interface, and Eq. (\ref{BL-force-balance0}) 
is verified.

Away from the fluid-solid interface and deep in the fluid, $g_x^{w0}$
and ${\sigma}_{zx}^0$ vanish while ${\sigma}_{xx}^0$ becomes a constant.
This leads to another integrated form for 
the static tangential force balance:
$$\begin{array}{ll}
 & \int_0^{z_d}dz[g_x^{w0}(x,z)+
\partial_x{\sigma}_{xx}^0(x,z)+\partial_z{\sigma}_{zx}^0(x,z)]\\
= & \int_0^{z_d}dz g_x^{w0}(x,z)+\partial_x\int_0^{z_d}dz{\sigma}^0_{xx}(x,z)\\
= & 0,
\end{array}$$
where $z_d$ represents a level $z=z_d$ far away from the interface
and deep in the fluid. Therefore, for the whole nonuniform fluid layer 
in equilibrium at the solid surface, there is no net tangential force 
(per unit area), although the fluid-solid interaction is modulated along
the tangential ($x$) direction. 
This is simply due to the cancellation of the wall force by the fluid force, 
because fluid molecules are adaptable to the modulated 
fluid-solid interaction.

\subsection{Dynamic Steady State}

\subsubsection{Hydrodynamic quantities}

In the static state, various forces and stresses show 
spatial variations in the molecular-scale vicinity of the wall.
The nonuniform static state forms a microscopic
background upon which hydrodynamic variations are generated 
when the fluid is driven (sheared) into hydrodynamic motion ($V\ne 0$). 
That's why the static quantities are taken as reference quantities.

As observed in MD simulations for $V\ne 0$, 
the variation of a particular static force/stress is still preserved
in the variation of the corresponding dynamic force/stress.
To obtain a hydrodynamic quantity, we need to subtract the static part
from the corresponding dynamic quantity. 
This subtraction scheme is formally expressed as
\begin{equation}\label{formal-subtraction}
\tilde{Q}=Q_{D}-Q_{S},
\end{equation}
where $Q_{S}$ is a static (reference) quantity,
$Q_{D}$ the corresponding dynamic quantity,
and the over tilde denotes the hydrodynamic quantity.
Physically, the characteristic variation magnitude of
$\tilde{Q}$ should be much smaller than that of 
$Q_{D}$ and $Q_{S}$.
In Figs. \ref{fig_BLforceshyd1} and \ref{fig_BLforceshyd2}, 
we show the profiles of
$\tilde{G}_x^{w}(x,\Delta z/2)$, 
$\int_0^{\Delta z}dz\tilde{\sigma}_{xx}(x,z)$, and
$\tilde{\sigma}_{zx}(x,\Delta z)$ for the boundary layer at the lower
fluid-solid interface. They are obtained according to 
Eq. (\ref{formal-subtraction}):
$$\tilde{G}_x^{w}(x,\Delta z/2)=
{G}_x^{w}(x,\Delta z/2)-{G}_x^{w0}(x,\Delta z/2),$$
$$\int_0^{\Delta z}dz\tilde{\sigma}_{xx}(x,z)=
\int_0^{\Delta z}dz\left[{\sigma}_{xx}(x,z)-{\sigma}_{xx}^0(x,z)\right],$$
and
$$\tilde{\sigma}_{zx}(x,\Delta z)=
{\sigma}_{zx}(x,\Delta z)-{\sigma}_{zx}^0(x,\Delta z).$$
It is readily observed that the variation of each hydrodynamic quantity
shows a magnitude much smaller than that seen from the corresponding
static quantity. This fact indicates that in deducing a continuum
relation (say, a slip boundary condition or a constitutive equation
for stress), if a hydrodynamic quantity ($\tilde{Q}$) is to be used, 
then it can by no means be replaced by a dynamic quantity ($Q_{D}$), 
without subtracting the static part. This will be shown in detail 
in Sec. \ref{slip-coefficient} for the Navier slip model and
Sec. \ref{newtonian-stress} for the Newtonian stress.

\subsubsection{Hydrodynamic force balance}

In the dynamic steady state ($V\ne 0$), local force balance still
requires the stress tangential to the fluid-solid interface to 
be the same on the two sides. 
An integrated form for the dynamic tangential force balance 
in the boundary layer at the lower fluid-solid interface is given by
\begin{equation}\label{BL-force-balancedyn}
G_x^{w}(x,\Delta z/2)+{G}_x^{f}(x,\Delta z/2)=0.
\end{equation}
Similar to $G_x^{w0}(x,\Delta z/2)$ and $G_x^{f0}(x,\Delta z/2)$,
$G_x^{w}(x,\Delta z/2)$ is the tangential wall force per unit area 
in this layer, given by
$$G_x^{w}(x,\Delta z/2)=\int_0^{\Delta z}dz g_x^{w}(x,z)$$ 
in terms of the tangential wall force density $g_x^{w}$,
and $G_x^{f}(x,\Delta z/2)$ is the tangential fluid force 
per unit area in the same layer, given by
\begin{equation}\label{fluid-force-expressiondyn}
\begin{array}{ll}
{G}_x^{f}(x,\Delta z/2) & =\int_0^{\Delta z}dz
[\partial_x{\sigma}_{xx}(x,z)+\partial_z{\sigma}_{zx}(x,z)]\\
& =\partial_x\int_0^{\Delta z}dz{\sigma}_{xx}(x,z)
+{\sigma}_{zx}(x,\Delta z),
\end{array}
\end{equation}
where ${\sigma}_{xx}$ and ${\sigma}_{zx}$ are the $xx$ and $zx$
components of fluid stress in the dynamic state. 
Again, the fact that $\sigma_{zx}(x,0)=0$ has been used in 
Eq. (\ref{fluid-force-expressiondyn}).
According to Eq. (\ref{formal-subtraction}), an integrated form for 
the hydrodynamic tangential force balance may be obtained
by subtracting Eq. (\ref{BL-force-balance0}) from 
Eq. (\ref{BL-force-balancedyn}):
\begin{equation}\label{BL-force-balancehyd}
\tilde{G}_x^{w}(x,\Delta z/2)+\tilde{G}_x^{f}(x,\Delta z/2)=0, 
\end{equation}
In Fig. \ref{fig_forcebalancehyd} we show
the profiles of $\tilde{G}_x^{w}$ and $-\tilde{G}_x^{f}$ for 
the boundary layer at the lower fluid-solid interface. It is seen
that Eq. (\ref{BL-force-balancehyd}) is verified.

Finally, we emphasize that the MD verification of 
Eqs. (\ref{BL-force-balance0}), (\ref{BL-force-balancedyn}), and 
(\ref{BL-force-balancehyd}) needs a reliable evaluation of 
the tangential fluid forces ${G}_x^{f0}$, ${G}_x^{f}$, and 
$\tilde{G}_x^{f}$. It is essential to use the expressions 
in Eqs. (\ref{fluid-force-expression0}) and
(\ref{fluid-force-expressiondyn}), together with a stress measurement 
scheme that is reliable near the fluid-solid interface
(see Appendix \ref{app-stress-measurement}).

\subsection{Boundary Layer and Sharp Boundary Limit}

From Eqs. (\ref{fluid-force-expression0}),
(\ref{fluid-force-expressiondyn}), and (\ref{formal-subtraction}),
we have the hydrodynamic tangential fluid force per unit area
\begin{equation}\label{fluid-force-expressionhyd}
\begin{array}{ll}
\tilde{G}_x^{f}(x,\Delta z/2) & =\int_0^{\Delta z}dz
[\partial_x\tilde{\sigma}_{xx}(x,z)+\partial_z\tilde{\sigma}_{zx}(x,z)]\\
& =\partial_x\int_0^{\Delta z}dz\tilde{\sigma}_{xx}(x,z)
+\tilde{\sigma}_{zx}(x,\Delta z),
\end{array}
\end{equation}
in the boundary layer at the lower fluid-solid interface.
Integrated forms for ${G}_x^{f0}$, ${G}_x^{f}$, and $\tilde{G}_x^{f}$
are necessary because the tangential wall force is distributed 
within a finite distance from the wall.
MD measurements show that, beyond the boundary layer 
(the first horizontal layer at the wall), 
the hydrodynamic tangential wall force $\tilde{G}_x^{w}$ vanishes. 
That is, the tangential wall force density $\tilde{g}_x^{w}$ 
is distributed within the boundary layer only, and hence 
$\tilde{G}_x^{w}(x,\Delta z/2)$ fully
accounts for the hydrodynamic tangential wall force 
(at the lower fluid-solid interface). The short-range nature of 
$\tilde{G}_x^{w}$ is due to the fact that as a hydrodynamic quantity
defined by Eq. (\ref{formal-subtraction}), $\tilde{G}_x^{w}$ solely 
arises from the ``roughness'' of wall potential through 
kinetic collisions of fluid molecules with wall molecules \cite{th-tro}. 
The modulation amplitude for this roughness decreases quickly with 
increasing distance from the wall. 
Hence the hydrodynamic tangential wall force tends to saturate at 
a relatively short range (see Appendix \ref{app-tangential-wall-force}).
Considering that $\tilde{G}_x^w$ exists in the boundary layer only, 
hereafter we use $\tilde{G}_x^w(x)$ ($=\tilde{G}_x^w(x,\Delta z/2)$)
to denote the total hydrodynamic tangential wall force per unit area.

We take the sharp boundary limit by letting $\tilde{G}_x^w$ 
strictly concentrate at $z=0$: 
$\tilde{g}_x^w(x,z)=\tilde{G}_x^w(x)\delta(z)$ with the same 
$\tilde{G}_x^w(x)$ per unit area. 
Rewriting $\tilde{G}_x^f$ in Eq. (\ref{fluid-force-expressionhyd}) as
$$\begin{array}{ll}
& \tilde{G}_x^f(x,\Delta z/2) \\ = & \int_{0^-}^{\Delta z}dz
[\partial_x\tilde{\sigma}_{xx}(x,z)+\partial_z\tilde{\sigma}_{zx}(x,z)]\\
= & \tilde{\sigma}_{zx}(x,0^+)+\int_{0^+}^{\Delta z}dz
[\partial_x\tilde{\sigma}_{xx}(x,z)+\partial_z\tilde{\sigma}_{zx}(x,z)],
\end{array}$$
we obtain 
$$\tilde{G}_x^f(x,\Delta z/2)=\tilde{\sigma}_{zx}(x,0^+),$$
because now local force balance requires $\partial_x\tilde{\sigma}_{xx}+
\partial_z\tilde{\sigma}_{zx}=0$ above $z=0^+$.
Hereafter we use $\tilde{G}_x^f(x)$  ($=\tilde{G}_x^f(x,\Delta z/2)$) 
to denote the hydrodynamic tangential fluid force per unit area 
in the boundary layer. Therefore,
\begin{equation}\label{fluid-force-expressionsharp}
\tilde{G}_x^f(x)=\tilde{\sigma}_{zx}(x,0^+)=
-\tilde{G}_x^w(x).
\end{equation}
In the sharp boundary limit, the boundary layer extends from 
$z=0^-$ to $z=0^+$, and $\tilde{\sigma}_{zx}$ varies from 
$\tilde{\sigma}_{zx}(x,0^-)=0$ to $\tilde{\sigma}_{zx}(x,0^+)=
\tilde{G}_x^f(x)$ at $z=0$ such that
$$(\nabla\cdot\tilde{\mbox{\boldmath$\sigma$}})\cdot\hat{\bf x}
=\tilde{G}_x^f\delta(z),$$
in balance with the tangential wall force density 
$\tilde{g}_x^w(x,z)=\tilde{G}_x^w(x)\delta(z)$. 
Equation. (\ref{fluid-force-expressionsharp}) may serve as 
a boundary condition in a continuum hydrodynamic model 
provided the continuum forms of 
$\tilde{\sigma}_{zx}(x,0^+)$ and $\tilde{G}_x^w(x)$ are given.
In Sec. \ref{slip-coefficient} we show the validity of the Navier slip model
for $\tilde{G}_x^w(x)$. In Sec. \ref{newtonian-stress} we show the validity
of the Newtonian constitutive relation for $\tilde{\sigma}_{zx}(x,0^+)$.

\section{TOWARDS A CONTINUUM HYDRODYNAMIC MODEL}\label{model}
\subsection{Local Slip Coefficient}\label{slip-coefficient}

Hydrodynamic motion of fluid at the fluid-solid interface
is represented by the slip velocity $v_x^{slip}$.
In MD simulations $v_x^{slip}$ is obtained from 
the tangential fluid velocity in the boundary layer, 
measured relative to the wall, 
i.e., $v_x^{slip}=v_x$ at the lower fluid-solid interface or 
$v_x^{slip}=v_x-V$ at the upper interface. 
The hydrodynamic viscous coupling between fluid and solid 
is described by a constitutive relation between 
the hydrodynamic tangential wall force $\tilde{G}_x^w$ and 
the slip velocity $v_x^{slip}$:
\begin{equation}\label{navier-wallforce}
\tilde{G}_x^w(x)=-\beta(x) v_x^{slip}(x),
\end{equation}
where $\beta(x)$ is the local slip coefficient at $x$. 
Equation (\ref{navier-wallforce}) is generally referred to as 
the Navier slip model. The validity of this model has been 
verified by many MD studies, yet most of them have been done 
for homogeneous solid surfaces only \cite{th-rob,th-tro,barrat,ckb}.
In the present problem, the local nature of the Navier slip model 
can be better revealed. Obviously, the slip coefficient $\beta$ 
must vary along the patterned surface where the fluid-solid interaction
is modulated. In particular, being a {\it local} coupling constant, 
$\beta$ should depend on $x$ through the $x$-dependent 
parameter $\delta_{fs}$ in potential $U_{fs}$, i.e., 
\begin{equation}\label{local-slipcoefficient}
\beta(x)=\beta(\delta_{fs}(x)),
\end{equation}
where $\beta$ as a function of $\delta_{fs}$,
denoted by $\beta(\delta_{fs})$, relates a microscopic interaction
parameter and a coefficient in the hydrodynamic slip model.
Equation (\ref{local-slipcoefficient}) reflects the short-range
nature of the fluid-solid interaction, and has been verified as follows.

A series of independent MD simulations have been carried out, for 
Couette flow between two identical homogeneous solid surfaces, using
two identical fluid-solid interaction potentials with a constant 
$\delta_{fs}$. Other simulation parameters remain unchanged, 
including $\rho$, $T$, $\rho_w$, $\epsilon_{fs}$, and $\sigma_{fs}$.
For each particular value of $\delta_{fs}$, the slip coefficient 
$\beta$ is obtained by measuring $\tilde{G}_x^w$ and $v_x^{slip}$, 
each being a constant along the $x$ direction.
The functional dependence of $\beta$ on $\delta_{fs}$, 
$\beta(\delta_{fs})$, has been numerically obtained 
by scanning a set of values for $\delta_{fs}$, 
as shown in Fig. \ref{fig_betavsdelta}. 
(This dependence is qualitatively consistent with the large slip effect 
at a nonwetting fluid-solid interface \cite{barrat}.)
Substituting into $\beta(\delta_{fs})$ the $\delta_{fs}(x)$ profile
in Eq. (\ref{periodic-delta}) then yields $\beta(x)$ 
in Eq. (\ref{local-slipcoefficient}) for the patterned surface. 
Using this $\beta(x)$, plus the slip velocity profile $v_x^{slip}(x)$
directly measured in the MD simulation (see Fig. \ref{fig_vxslip}),
we calculate the hydrodynamic tangential wall force 
at the patterned surface according to Eq. (\ref{navier-wallforce}). 
This calculated $\tilde{G}_x^w(x)$ is shown in
Fig. \ref{fig_wallforcecomparison}, and
compared to the measured $\tilde{G}_x^w$ 
(which has already been given in Fig. \ref{fig_BLforceshyd1}). 
The good agreement clearly verifies the local Navier slip model expressed 
by Eqs. (\ref{navier-wallforce}) and (\ref{local-slipcoefficient}).
Of course, the local dependence of $\beta$ on $\delta_{fs}$, 
expressed by Eq. (\ref{local-slipcoefficient}), is not without limit. 
Physically, $\beta$ is defined as a phenomenological parameter over
a microscopic length scale that is about a few lattice constants of 
the atomistic wall. Therefore, 
to validate Eq. (\ref{local-slipcoefficient}) for $\beta(x)$ in
Eq. (\ref{navier-wallforce}), the lateral variation of 
$\delta_{fs}$ must be slow. That is, the characteristic distance
over which  $\delta_{fs}$ varies must be much larger compared to
the wall lattice constant.
That's why in Eq. (\ref{periodic-delta}) the modulation period $P$ 
is chosen to be larger than $10\sigma$.  
In fact, the disagreement between MD and continuum results has already
been observed when the period of surface pattern approaches 
a molecular scale \cite{troian-preprint}.

\subsection{Newtonian Stress}\label{newtonian-stress}

To obtain a continuum hydrodynamic model for fluid flow past 
a chemically patterned solid surface, we need a momentum transport
equation and a slip boundary condition.
Far away from the solid surface, the molecular density is uniform, 
the (hydrodynamic) stress is Newtonian, and the momentum transport 
can be well described by the Navier-Stokes equation
\begin{equation}\label{model-NSE}
\rho\left[\displaystyle\frac{\partial{\bf v}}{\partial t}+
\left({\bf v}\cdot\nabla\right){\bf v}\right]=
-\nabla p+\eta\nabla^2{\bf v},
\end{equation}
where the viscosity $\eta=2.1\sqrt{\epsilon m}/\sigma^2$
has been determined by MD measurements. 
Close to the rigid wall, however, the fluid density shows 
a short-range oscillation along the fluid-solid interface normal 
($z$) across a few molecular layers \cite{israelachivili}. 
It has been shown that, in the presence of 
this normal density variation, MD flow fields still can be well 
reproduced by continuum calculations using a constant viscosity, 
unless the channel is extremely narrow 
($H\approx 4\sigma$) \cite{subcontinuum1,subcontinuum2}.

In the present problem, the fluid density shows a lateral oscillation
(along $x$) as well, imposed by the patterned solid surface.
With the lateral and normal density variations both present
near the solid surface, we proceed by using a constant viscosity 
throughout the fluid space in our continuum model. 
This is based on the observation that the magnitude 
of lateral density variation is on the same order of 
that of normal variation. 
To see the validity of the constant viscosity approximation, 
the hydrodynamic tangential stress $\tilde{\sigma}_{zx}(x,z)$ 
has been measured at $z=2\Delta z$ where density oscillation
is $\approx 10\%$. This measured quantity is then compared to that
calculated from the Newtonian relation
$\eta (\partial_zv_x+\partial_xv_z)$ using the MD measured
velocity field (see Fig. \ref{fig_newton}). 
The overall agreement supports the use of a constant viscosity.

The hydrodynamic slip boundary condition is obtained by combining
Eqs. (\ref{fluid-force-expressionsharp}), (\ref{navier-wallforce}), 
and (\ref{local-slipcoefficient}) with
the Newtonian constitutive relation for the tangential stress
$\tilde{\sigma}_{zx}$ at the solid surface. This leads to
the Navier slip boundary condition 
\begin{equation}\label{model-NBC}
\beta(\delta_{fs}(x))v_x^{slip}=-\eta \partial_n v_x,
\end{equation}
with a local slip coefficient $\beta(\delta_{fs}(x))$
(here $n$ denotes the outward surface normal, 
$-z$ at the lower surface or $z$ at the upper surface).
A local slip length, $l_s(x)$, can be defined by
$l_s(x)=\eta/\beta(\delta_{fs}(x))$.

\begin{figure}[h]
\centerline{\psfig{figure=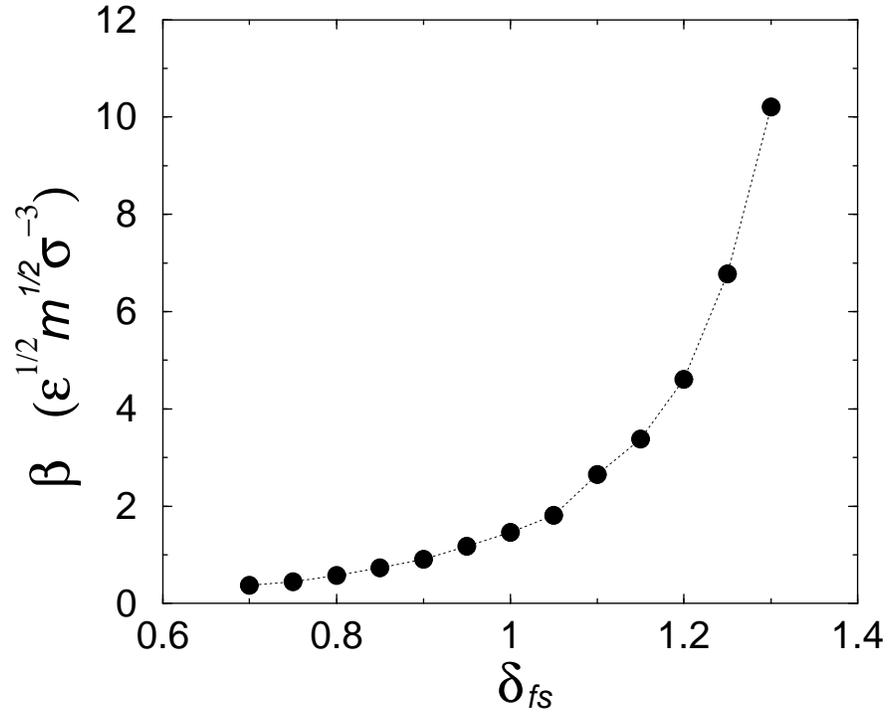,height=10cm}}
\bigskip
\caption{\small Slip coefficient $\beta$ plotted as a function of
the parameter $\delta_{fs}$ in $U_{fs}(r)$. 
}\label{fig_betavsdelta}
\end{figure}

To summarize, our hydrodynamic model consists of the Navier-Stokes equation (\ref{model-NSE}), the NBC (\ref{model-NBC}),
the impermeability boundary condition $v_n=0$,
and the incompressibility condition $\nabla\cdot{\bf v}=0$.
Continuum calculations involve six parameters, including 
the system dimensions $L$ along $x$ and $H$ along $z$, 
the shearing speed $V$, the fluid density $\rho$, the viscosity $\eta$,
and the slip coefficient $\beta$ (as a function of $\delta_{fs}$). 
We emphasize that our model has {\it no} adjustable parameter:  
the values for $L$, $H$, $V$, $\rho$ come directly from MD simulations,
$\eta$ has been determined to be $2.1\sqrt{\epsilon m}/\sigma^2$,
and $\beta$ has been measured as a function of $\delta_{fs}$ 
(see Fig. \ref{fig_betavsdelta}).
A continuous dependence of $\beta$ on $\delta_{fs}$ can be
obtained from the MD data (in Fig. \ref{fig_betavsdelta})
through interpolation.

\section{CONTINUUM HYDRODYNAMICS RESULTS}\label{continuum}

A second order finite-difference scheme is designed to solve 
the Navier-Stokes equation. Essentially, it is a modified version of 
the pressure-Poisson formulation \cite{liu}, 
where the incompressibility condition is replaced by 
the pressure Poisson equation 
plus a divergence-free boundary condition for the velocity. 
The pressure-Poisson formulation has already been used to solve 
the Navier-Stokes equation for immiscible two-phase flow 
in the vicinity of the moving contact line, where a generalized NBC 
is involved \cite{qws-prl,qws-pre}. (For details of 
the numerical algorithm, see Appendix C of Ref. \cite{qws-pre}. 
It can be easily simplified to work for the present single-phase problem.)

\subsection{Comparison of MD and Continuum Results}\label{comparison}

The two equations of our continuum model, Eqs. (\ref{model-NSE}) 
and (\ref{model-NBC}), come directly from the MD verification of 
the Newtonian stress (see Fig. \ref{fig_newton}) and 
the local NBC (see Eqs. (\ref{navier-wallforce}) and
(\ref{local-slipcoefficient}), plus Fig. \ref{fig_wallforcecomparison}).
Therefore, predictions from the continuum model are expected to 
match MD results quantitatively.

\begin{figure}[h]
\centerline{\psfig{figure=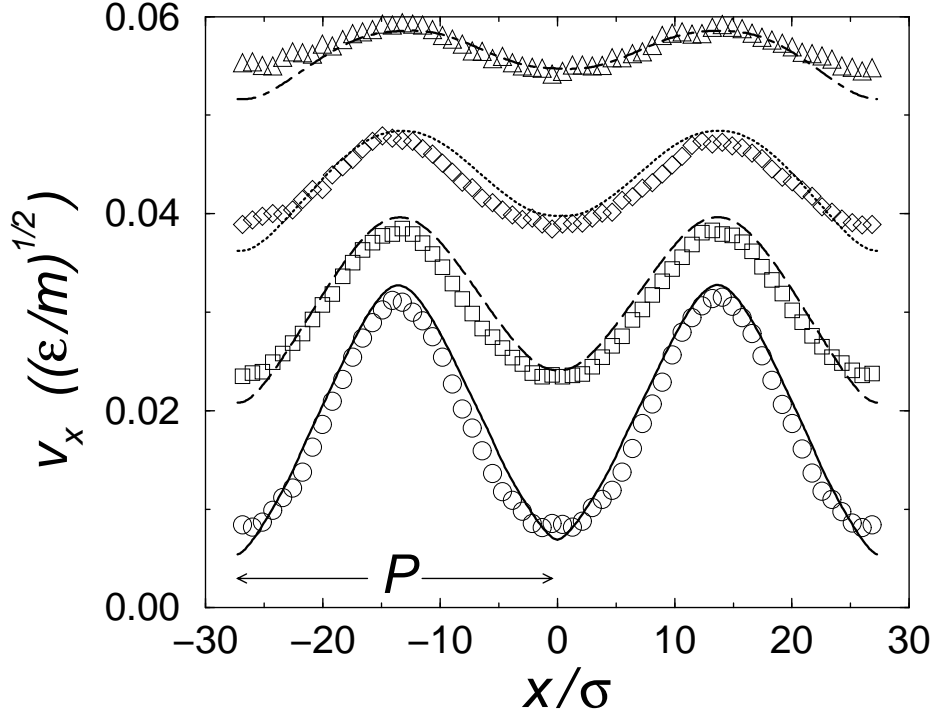,height=10cm}}
\bigskip
\caption{\small $v_x$ plotted as a function of $x$ for four $z$ levels
close to the lower wall of sinusoidal fluid-solid interaction
(see Eq. (\ref{periodic-delta}) for $\delta_{fs}(x)$ of period $P$). 
The symbols denote the MD data and the lines
represent the corresponding continuum results. Here the MD data are
obtained for a system of $H=13.6\sigma$, $L=2P=54.4\sigma$, and 
$V=0.25\sqrt{\epsilon/m}$.
The four $z$ levels are at $z=0.425\sigma$ (circles and solid line), 
$1.275\sigma$ (squares and dashed line), 
$2.125\sigma$ (diamonds and dotted line), 
and $2.975\sigma$ (triangles and dash-dotted line).
}\label{fig_vx_osci}
\end{figure}

Figure \ref{fig_vx_osci} shows the periodic $v_x$ profiles close to 
the lower wall of sinusoidal periodic fluid-solid interaction.

\begin{figure}[h]
\centerline{\psfig{figure=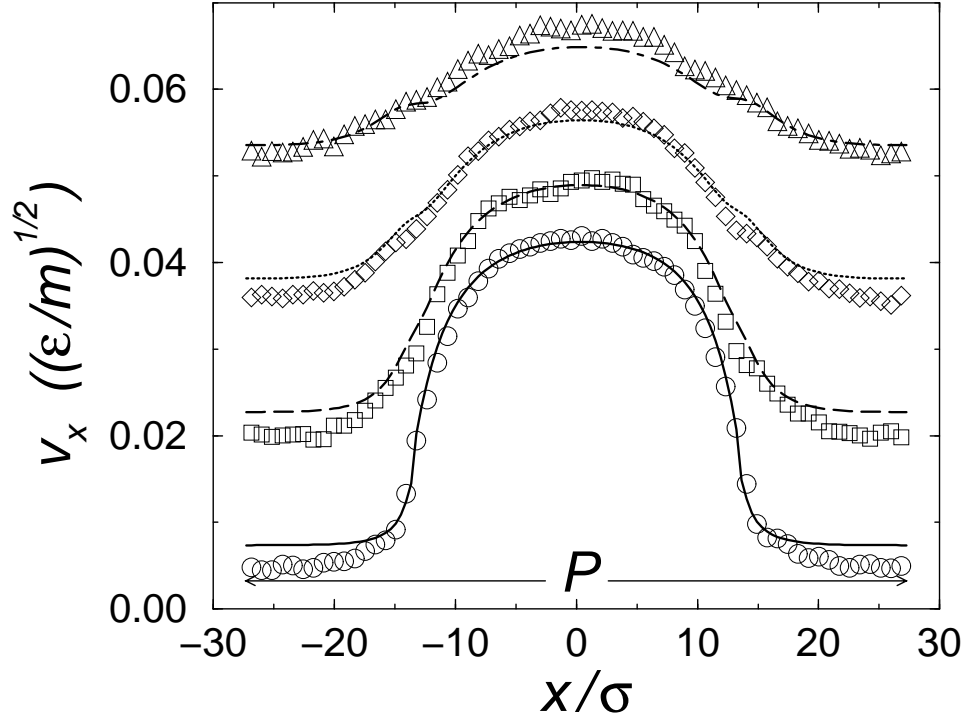,height=10cm}}
\bigskip
\caption{\small $v_x$ plotted as a function of $x$ for four $z$ levels
close to the lower wall of stepwise fluid-solid interaction.
The symbols denote the MD data and the lines
represent the corresponding continuum results. Here the MD data are
obtained for a system of $H=13.6\sigma$, $L=P=54.4\sigma$, and 
$V=0.25\sqrt{\epsilon/m}$.
The four $z$ levels are at $z=0.425\sigma$ (circles and solid line), 
$1.275\sigma$ (squares and dashed line), 
$2.125\sigma$ (diamonds and dotted line), 
and $2.975\sigma$ (triangles and dash-dotted line).
}\label{fig_vx_step}
\end{figure}

Figure \ref{fig_vx_step} shows the periodic $v_x$ profiles close to 
the lower wall of stepwise periodic fluid-solid interaction. 
(This corresponds to the transverse case studied in 
Ref. \cite{troian-preprint}.)
In the first period $-P/2\le x<P/2$, 
$\delta_{fs}(x)=\delta_0-\delta_1$ for $-P/4\le x<P/4$
or $\delta_0+\delta_1$ elsewhere.

The overall agreement in the above two cases affirms the validity of
the local NBC and the hydrodynamic model. The small discrepancies 
seen in Figs. \ref{fig_vx_osci} and \ref{fig_vx_step} may be attributed
to the relatively small $H$ ($=13.6\sigma$), the nonuniform 
boundary-layer fluid, and the relatively fast variation of $\delta_{fs}$
(especially in the sinusoidal case).

\subsection{Adjustable Slip Length}\label{adjust}

Consider a chemically patterned surface composed of a periodic array 
of stripes parallel to the $y$ axis.
Each stripe is of type $A$ or $B$, and the periodicity along the $x$
direction is realized by arranging the stripes according to 
$\cdot\cdot\cdot ABABAB\cdot\cdot\cdot$ (see Fig. \ref{fig_pattern}).
This way of patterning has been modeled in MD simulations
by a stepwise modulation of fluid-solid interaction, and 
the MD and continuum results agree well (see Fig. \ref{fig_vx_step}).

The continuum model uses two different slip lengths $l_{sA}$ and $l_{sB}$
for the $A$ and $B$ stripes, respectively. 
While the chemical property of each stripe is kept invariant, 
the surface pattern may be continuously changed by varying 
the stripe widths. With the two slip lengths fixed, continuum flow fields 
have been calculated for stripe widths varied in a wide range. 
It is found that the slip behavior can continuously change from 
the regime of distinct slip domains (for large stripe widths) 
to the regime of effective medium (for small stripe widths, 
with an effective slip length showing up). 
Note that in the latter regime, the stripe widths, though small 
compared to the slip length, are still considered large enough to justify 
the validity of the continuum model. The effective-medium regime 
provides a way to systematically tune the effective slip length. 
A desired slip length, if not naturally produced at a chemically
homogeneous surface, can thus be realized at a patterned surface.

Continuum hydrodynamic calculations have been carried out for Couette flow
between two solid planar surfaces parallel to the $xy$ plane.
The lower surface is patterned and the upper surface is homogeneous.
The fluid is sheared by fixing the lower surface and translating
the upper surface at a constant velocity $V$ in the $x$ direction.
We focus on the regime of viscous flow with small Reynolds numbers 
($Re=\rho VH/\eta\approx 1$, where $H$ is the distance between 
the two solid surfaces). With the fluid velocity $\bf v$ measured by
$V$ and the coordinates $x$ and $z$ measured by $H$,
the scaled steady-state velocity fields, ${\bf v}(x/H,z/H)/V$,
are controlled by the dimensionless parameters
$l_{sA}/H$, $l_{sB}/H$, $w_A/H$, and $w_B/H$, where $w_A$ and $w_B$
are the widths of $A$ and $B$ stripes, respectively.

\begin{figure}[h]
\centerline{\psfig{figure=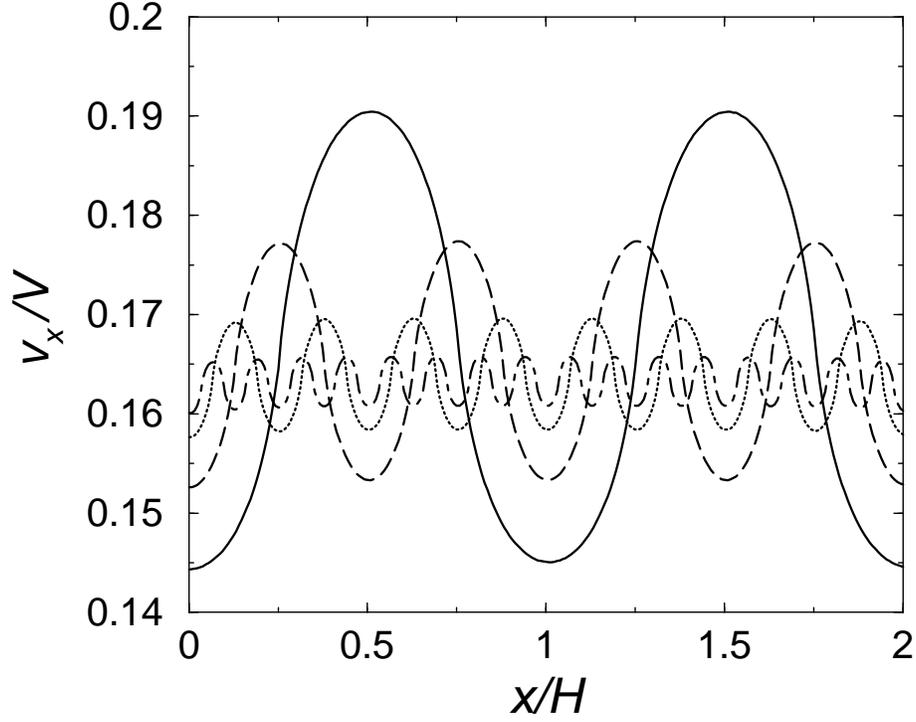,height=10cm}}
\bigskip
\caption{\small Slip profiles at differently patterned solid surfaces.
Here the scaled slip velocity $v_x/V$ is plotted as a function 
of $x/H$. The pattern period $P=w_A+w_B$ varies from $H$ to $H/8$, 
with equal stripe widths ($w_A=w_B$). The solid line is for $P/H=1$, 
the dashed line for $P/H=1/2$, the dotted line for $P/H=1/4$, 
and the dash-dotted line for $P/H=1/8$.
}\label{fig_slip_pattern1}
\end{figure}

\begin{figure}[h]
\centerline{\psfig{figure=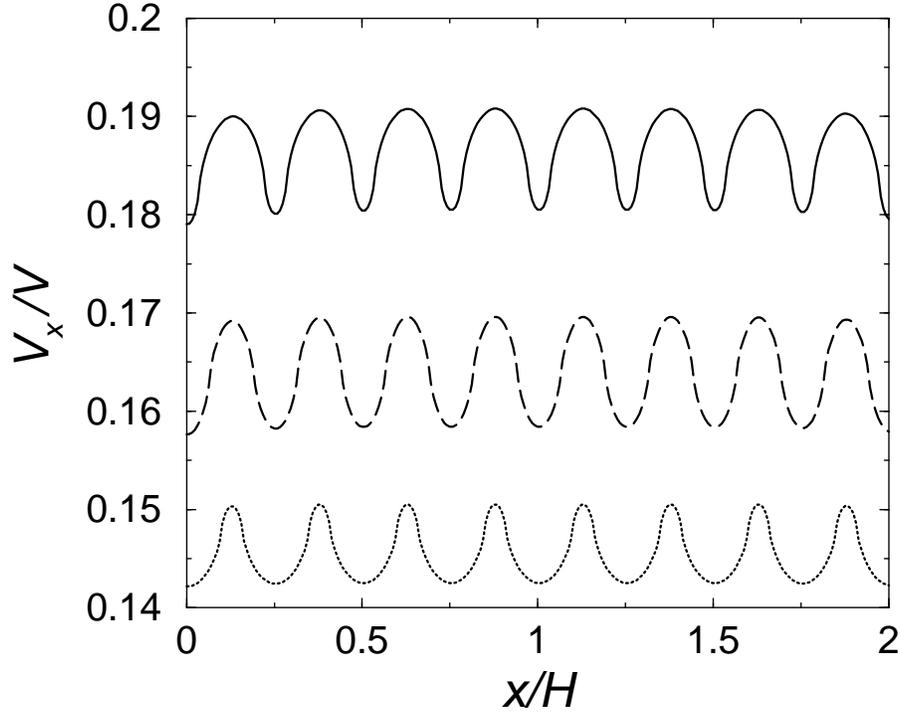,height=10cm}}
\bigskip
\caption{\small Slip profiles at differently patterned solid surfaces.
Here the scaled slip velocity $v_x/V$ is plotted as a function 
of $x/H$. The pattern period $P=w_A+w_B$ equals $H/4$, with 
$w_A/w_B$ varying from $1/3$ to $3$. 
The solid line is for $w_A/w_B=1/3$, the dashed line for $w_A/w_B=1$
(same as the dotted line in Fig. \ref{fig_slip_pattern1}),
and the dotted line for $w_A/w_B=3$.
}\label{fig_slip_pattern2}
\end{figure}

Figures \ref{fig_slip_pattern1} and \ref{fig_slip_pattern2}
show the steady-state slip profiles at six differently patterned surfaces, 
obtained for $l_{sA}/H=0.147$ and $l_{sB}/H=0.441$.
The values of $w_A/H$ and $w_B/H$ have been varied to show 
the control of an effective slip length by surface patterning.
In the present Couette-flow geometry, the slip velocity has to oscillate 
in the $x$ direction at the lower surface, where the slip length
is modeled as a periodic function of $x$ according to the surface patterning.
From Fig. \ref{fig_slip_pattern1}, it is obvious that the slip velocity 
becomes less oscillatory in magnitude when the pattern period $P$ is decreased. 
In fact, for $P/H=1/4$ and $1/8$, the relative magnitude of
oscillation for $v_x$, defined as $(v_x^{max}-v_x^{min})/2v_x^{ave}$,
becomes $3.3\%$ and $1.5\%$, respectively, while for $P/H=1$, 
that oscillation magnitude is $14\%$.
Therefore, when the solid surface is patterned with a sufficiently small
period, it appears to be homogeneous. The nearly uniform slip velocity
can be used to define an effective slip length $l_s^{eff}$. 
For the present Couette flow, it is given by 
$$\displaystyle\frac{l_s^{eff}}{H}=\displaystyle\frac{v_x^{ave}}
{V-v_x^{ave}},$$
which yields $l_s^{eff}/H=0.196$ from the data in Fig. \ref{fig_slip_pattern1}.
For sufficiently small pattern period $P$, the effective slip length
can be further tuned by varying the ratio of the $A$ stripes 
(with a small slip length) to the $B$ stripes (with a large slip length).
Figure \ref{fig_slip_pattern2} shows that by fixing $P/H$ at $1/4$ and
increasing the proportion of the $A$ stripes, the average amount of slip
is appreciably reduced. That is, the effective slip length $l_s^{eff}$
decreases with the increasing proportion of the $A$ stripes.

\section{CONCLUSION AND DISCUSSION}\label{disc}

In this paper we have used both the MD and continuum simulations to 
study the slip boundary condition for single-phase flow past 
a chemically patterned surface. Modulation of fluid-solid interaction 
along the patterned surface induces a lateral variation in various 
stresses and forces in the fluid. This has been observed in both 
the static equilibrium state and the dynamic steady state. Given this
lateral variation, a subtraction scheme has been developed for extracting 
hydrodynamic quantities from MD simulations. Based on this scheme, 
the validity of the Navier boundary condition has been verified in 
describing the local hydrodynamic viscous coupling between the fluid 
and the patterned solid surface. A continuum hydrodynamic model has 
been formulated. It is capable of yielding steady-state flow fields
in quantitative agreement with those from MD simulations.
We have used the continuum model to show that an effective slip length
can be realized and tuned by surface patterning.

The framework of heterogeneous multiscale method \cite{hmm} 
provides a general scheme for developing hybrid 
atomistic-continuum methods. The hybrid methods use MD to 
extract boundary condition(s) and/or constitutive relation(s) 
for continuum calculation. Whatever a continuum calculation needs, 
the relevant hydrodynamic quantities have to be properly extracted 
from MD data. In the present problem, we see that a subtraction 
of the static component from the dynamic quantity is essential to 
obtaining the correct hydrodynamic information.
We emphasize that hybrid methods are mostly used to 
investigate problems involving complex fluids, 
complex fluid-solid interactions, and/or complex geometries, 
while in such problems, the static configurations are usually 
nontrivial in a molecular point of view.

\section*{Acknowledgments}

Partial support from Hong Kong RGC DAG 03/04.SC21 is hereby acknowledged.
We would like to thank Prof. Weinan E and Dr. Weiqing Ren for 
helpful discussion.

\appendix
\section{Atomistic Solid Walls}
\label{app-atomistic-solid-walls}

The MD results presented in this paper were obtained from simulations
using solid walls constructed by two [001] planes of an fcc lattice.
We have also carried out MD simulations using thicker confining walls.
The number of crystalline layers ([001] planes of fcc lattice)
is changed from two to four in constructing each of the two walls. 
The fluid-solid
interaction potential $U_{fs}$ is then cut off at a larger distance
$r_c=2.5\sigma$. It turned out that both the velocity field and
the stress field remain essentially the same.

\section{Tangential Wall Force}
\label{app-tangential-wall-force}

In terms of the tangential wall force density $g_x^w$, we define 
the tangential wall force per unit area in a horizontal layer as 
$G_x^w(x,\frac{1}{2}(z_1+z_2))=\int_{z_1}^{z_2}dz {g}_x^w(x,z)$, 
which is the total tangential wall force accumulated across that layer,
sandwiched between $z=z_1$ and $z=z_2$.   

The wall force can be singled out by separating the force on 
each fluid molecule into fluid-solid and fluid-fluid components. 
The fluid molecules close to the crystalline wall experience 
a {\it fast} periodic modulation in interaction with the wall.
When coupled with kinetic collisions with the wall molecules, 
there arises the first component in the tangential wall force, 
sharply distributed along $z$ close to the wall.
In the presence of a periodically varied fluid-solid interaction 
(at the lower fluid-solid interface, with a period much larger than 
that of the crystalline structure), the second component in 
the tangential wall force arises along with 
a {\it slow} oscillation of fluid molecules' distribution. 
This component is also narrowly distributed along $z$ near the wall, 
but with a distribution width larger than that of the first component.
This may be understood as follows. 

Two quite different periods exist in the periodic modulation 
in fluid molecules' interaction with the wall. 
The first period is from the crystalline structure of the wall. 
For a fluid molecule close to the solid wall,
the interaction with one particular (the closest) wall molecule 
can be much stronger than all the others. As this fluid molecule moves
laterally but remaining close to the wall, it would thus experience 
a strong periodic modulation in its interaction with the wall.
This lateral inhomogeneity is generally referred to as 
the ``roughness'' of the wall potential \cite{th-tro}. 
Away from the fluid-solid interface, each fluid molecule can interact 
with many wall molecules on a nearly equal basis. 
Thus the modulation amplitude of the wall potential would clearly 
decrease with increasing distance from the wall. 
Hence the first component of the tangential wall force 
tends to saturate at a relatively short range.
The second period, being much larger than the first one, 
comes directly from the fluid-solid interaction potential. As a consequence, 
the distribution of fluid molecules would be periodically adjusted  
along the wall. This occurs within a relatively long distance
from the surface, 
as determined by the range of fluid-solid interaction and 
the correlation length of fluid. Therefore, the second component of 
the tangential wall force has a wider distribution than the first one.

\section{Stress Measurement}
\label{app-stress-measurement}

The Irving-Kirkwood expression \cite{irving-kirkwood} has been widely 
used for stress measurement in MD simulations. However, as pointed out 
by the authors themselves, the leading-order expression 
for the interaction contribution to the stress tensor
represents only the leading term in an asymptotic expansion, 
accurate when the interaction range is small compared to the range of 
hydrodynamic variation. As a consequence, this leading-order expression 
is not reliable near a fluid-fluid or a fluid-solid interface. 
In this paper, the stress components $\sigma_{xx}$ and $\sigma_{zx}$ 
are obtained from the time averages of the kinetic momentum transfer 
plus the fluid-fluid interaction forces across the fixed $x$ and $z$ 
bin surfaces. 
In particular, we directly measure the $x$ component of 
the fluid-fluid interaction forces acting across the $x(z)$ 
bin surfaces, in order to obtain the $xx(zx)$ component of 
${\mbox{\boldmath$\sigma$}}_U$, the interaction contribution to 
the stress tensor ${\mbox{\boldmath$\sigma$}}$. 
Here ``acting across'' means that the line 
connecting a pair of molecules intersects the bin surface 
(the so-called Irving-Kirkwood convention \cite{irving-kirkwood}). 
For more details, see Appendix B in Ref. \cite{qws-pre}.

\newpage

\newpage

\newpage

\newpage

\newpage

\newpage

\newpage

\newpage

\newpage

\newpage

\newpage

\newpage

\newpage

\newpage

\newpage

\newpage

\end{document}